\documentstyle[11pt,newpasp,twoside,epsf]{article}
\def\edcomment#1{\iffalse\marginpar{\raggedright\sl#1\/}\else\relax\fi}
\reversemarginpar

\begin{document}
\title{The Galactic Halo and CDM}
\author{Michael R. Merrifield}
\affil{School of Physics \& Astronomy, University of Nottingham,
       University Park, Nottingham, NG7 2RD, UK}

\begin{abstract}
This paper reviews the available information on the central density
distribution and shape of the Milky Way's halo.  At present, there is
no strong evidence that the Milky Way's halo properties conflict with
the predictions of cold dark matter (CDM): a primordial central power
law cusp can be accommodated by the observations, and the current
constraints on flattening are also consistent with the predictions of
the theory.  If you want to pick a fight with CDM, then the Milky Way is
probably not the place to do it.
\end{abstract}

\section{Introduction}
The cosmological principle states that there are no special places in
the Universe.  A simple corollary of this principle is that we cannot
be anywhere special in the Universe, and this assumption has been
supported by a long line of discoveries running all the way from
Copernicus to Shapley demonstrating that we live in the sprawling
suburbs of the Milky Way.  It also, of course, means that the Milky
Way is a very ordinary spiral galaxy.  Although we should not be
fooled into thinking that all galaxies are necessarily like the Milky
Way, we have a very useful test of any theory of galaxy formation in that
properties of the Milky Way should lie within the range of what is
considered ``normal'' by the theory.

In some ways, we are making life hard for ourselves by using the Milky
Way for such tests, since many observations that are fairly trivial in
external galaxies are rather challenging in our own galaxy.  Our
position within the Milky Way complicates the issue for several
reasons.  First, the geometry is more complex than for distant
systems, since there is no simple relationship between angular and
linear scale.  Second, since parts of the Milky Way lie in all
directions, large surveys are generally required to study its
properties.  Third, living right in the dust lane of the Galaxy makes
obscuration more of a factor than in most external systems.  However,
these difficulties mainly affect attempts to measure global properties
of the Milky Way, such as its overall morphology.  For smaller-scale
phenomena, our proximity to the action makes the Galaxy an ideal
laboratory.

In particular, the Milky Way offers the unique possibility for an {\it
in situ} local measurement of halo properties that is not attainable
for any other system.  By measuring local stellar kinematics, Kuijken
\& Gilmore (1991) obtained a fairly robust estimate for the total
amount of mass in the solar neighborhood within $1.1\,{\rm kpc}$ of
the Galactic plane, $\Sigma_{1.1} \approx 70M_\odot\,{\rm pc}^{-2}$.
Much of this mass can be attributed to the visible components of the
Galaxy: $\sim 25M_\odot\,{\rm pc}^{-2}$ is contributed by the normal
stellar component, and $\sim 15M_\odot\,{\rm pc}^{-2}$ comes from the
interstellar medium (Olling \& Merrifield 2001).  However, this still
leaves around $30M_\odot\,{\rm pc}^{-2}$ unaccounted for, which
presumably must be attributed to the dark matter halo.  If the
scaleheight of this dark component is large compared to the $1.1\,{\rm
kpc}$ within which the mass is measured, then we can obtain a direct
measure of the local dark matter density, $\rho^{\rm DM}(R_0, 0)
\approx 0.014M_\odot\,{\rm pc}^{-3}$.  Here, we have made explicit the
Galactocentric cylindrical polar coordinates which locate the Sun at a
radius $R_0 \approx 8.5\,{\rm kpc}$ from the Galactic center (Kerr \&
Lynden-Bell 1986), approximately in the $z=0$ plane.  As we shall see
below, this single data point is probably the most important
contribution that the Milky Way can make to the general study of dark
matter, since it offers a unique localized measurement that we cannot
make at any other point, or in any other dark matter halo.

\section{Central Cusp}\label{sec:cusp}
One immediate constraint that the local dark matter measurement offers
is on the power law slope of any central cusp in the dark matter
distribution.  Almost all simulations indicate that cold dark matter
(CDM) should form halos with a density distribution that goes from a
relatively shallow power law at small radii to a steeper one at large
radii, with a transition occuring at a radius $r_s$ that depends on
the mass of the halo.  For an object the size of the Milky Way, this
transition should occur at $r_s \sim 20\,{\rm kpc}$, well outside
$R_0$, so the density distribution in the inner Galaxy should follow a
single power law rather closely.  

If we write this power law as $\rho(r) \propto r^{-\alpha}$ (where we
have assumed, for the moment, that the halo is spherical), and
normalize the distribution using $\rho^{\rm DM}(R_0, 0)$, we find that
the total mass of dark matter interior to the Solar radius is
\begin{equation} \label{eq:MDM}
M^{\rm DM}(R_0) = {4\pi \over 3 - \alpha} \rho^{\rm DM}(R_0, 0) \times R_0^3
\approx {1.0 \times 10^{11} \over 3 - \alpha} M_\odot.
\end{equation}

The total amount of baryonic material (stars plus interstellar medium)
in the inner Galaxy can be estimated by modeling these components, and
their gravitational influence can be calculated.  After conducting
such a census, Olling \& Merrifield (2001) concluded that this
material has the same gravitational influence as a spherical
distribution of mass $M^{\rm baryon}(R_0) \approx 5.7 \times 10^{10}
M_\odot$.\footnote{This is not quite the same as the total mass of
baryonic material, since this material is not in reality distributed
spherically.}

These masses can be compared to that which is inferred dynamically by
equating the centripetal acceleration of an object on a circular orbit
around the Galaxy at $R_0$ to the acceleration induced by gravity at
this radius.  This value depends on the rate at which an object would
have to travel to follow a circular orbit of radius $R_0$ around the
Milky Way, $\Theta_0 \approx 220\,{\rm km}\,{\rm s}^{-1}$ (Kerr \&
Lynden-Bell 1986), which gives a dynamical mass of 
\begin{equation}\label{eq:Mdyn}
M^{\rm total}(R_0) = {R_0 \Theta_0^2 \over G}
                   \approx 9.5 \times 10^{10} M_\odot. 
\end{equation}

Subtracting $M^{\rm baryon}(R_0)$ from this total mass gives a measure
of the contribution from dark matter, which can be equated to the
value in Equation~\ref{eq:MDM}.  Thus, we can solve for the only
unknown in Equation~\ref{eq:MDM}, the power law index $\alpha$, from
which we find a value of $\alpha \approx 0.4$.

Binney \& Evans (2001) performed a similar but much more thorough
analysis in which they looked at the enclosed mass at all radii in the
galaxy (by fitting to the full rotation curve rather than just the
local value of the circular velocity at $R_0$).  They also used the
microlensing optical depth toward the Galactic center to constrain the
baryon density in the inner Galaxy.  Since the microlensing
measurements place a similar amount of mass in baryons, $M^{\mu{\rm
lens}}(R_0)\approx 4 \times 10^{10} M_\odot$, and the strongest limits
on the halo mass come from the circular speed near the Solar radius in
the Galaxy, they reached a similar conclusion that $\alpha \la 0.4$.  

There are also other lines of evidence that the dark matter
distribution in the middle of the Milky Way cannot be strongly
centrally concentrated.  In particular, the Milky Way is known to be a
barred galaxy, and the rate at which its central bar rotates tells us
something about the distribution of dark matter in the vicinity.
Modeling of the HI and CO kinematics near the Galactic center
(Englmaier \& Gerhard 1999), associating features in the stellar
distribution of the Galaxy with orbital resonances (Sevenster 1999)
and Lagrange points (Binney, Gerhard \& Spergel 1997), and direct
measurements of the pattern speed (Debattista, Gerhard \& Sevenster
2002) all indicate that the bar is rotating rapidly.  In this context,
``rapid'' means that the bar pattern is rotating at such a speed that it
travels at approximately the same speed that stars near the end of the
bar orbit the Galaxy.  This is as fast as a bar pattern can rotate, but
there is no reason why it has to travel so quickly.  Indeed, if a bar
is embedded in a massive dark halo, then simulations have shown that
the transfer of angular momentum from the bar to the dark halo should
rapidly slow the bar from this maximum speed (Debattista \& Sellwood
2000).  Therefore, the rapid observed rotation speed of the Galactic
bar provides evidence that the Milky Way cannot have a lot of dark
matter near its center to absorb the bar's angular momentum.  This
argument implies that any central cusp in the dark matter distribution
must be rather weak, in line with the value of $\alpha \approx 0.4$
derived above.

This low value for the central power law cusp in the Milky Way is not
what the CDM simulations predict.  There has been some disagreement
amongst the simulators as to the exact value of $\alpha$, with a
lengthy argument as to whether it lies closer to 1.0 (Navarro, Frenk \&
White 1996) or 1.5 (Moore et al.\ 1999).  However, one thing upon which all
would agree is that the central cusp of a CDM halo should
have a power law index much larger than the value of 0.4 that we have
inferred for the Milky Way.  Thus, we would appear to have dealt a
killer blow to CDM: in the one place that this type of measurement can
be made reliably, the observations conflict grossly with the
predictions of the theory.

However, the case for the prosecution is not quite as convincing as it
might at first appear.  First of all, even if a galaxy initially
formed with a steep power law cusp in its dark halo as CDM requires,
there are mechanisms that could have erased it by the present day.  In
particular, as described above, the interaction between a bar and a
central cusp can redistribute angular momentum rather effectively, so
a strong bar could fling dark matter out from the center of a galaxy,
rearranging any initial cusp into a less concentrated distribution.
Unfortunately, once again the theorists are unable to agree on the
efficacy of this mechanism: Weinberg \& Katz (2002) have shown that a
strong initial bar in the Milky Way could have wiped out a central
cusp, whereas Sellwood (2003) has argued that any plausible bar would
lack sufficient angular momentum to do anything significant to the
dark matter distribution.  It is worth noting, though, that the above
simple calculation on the Milky Way implies that any redistribution of
mass would have to be pretty radical: since our calculation
effectively measures the total mass of dark matter inside $R_0$ in the
Milky Way, the deficit from a primordial $\alpha \approx 1$ cusp could
only be explained if the initial cusp material had been flung all the
way out to beyond $R_0$.

A second weakness in the above argument for a low value of $\alpha$
is its reliance on various Galactic constants.  There are
significant uncertainties in the contributions of baryonic material to
the local density in the Galaxy, which translate directly into
uncertainties in the amount of the total density that must be
attributed to dark matter.  Further, The total dynamical mass
constraint depends on the adopted values for the basic Galactic
constants $R_0$ and $\Theta_0$ (see Equation~\ref{eq:Mdyn}).  Thus,
for example, Binney \& Evans' (2001) exclusion of an $\alpha = 1$
model depends largely on the fact that such a model would require the
local rotation speed to be $\approx 230\,{\rm km}\,{\rm s}^{-1}$, in
conflict with their adopted conventional value of $\Theta_0 =
220\,{\rm km}\,{\rm s}^{-1}$.  Although it might be considered
inappropriate to suggest such a thing at an IAU-sponsored symposium,
it is likely that the IAU-sanctioned ``official'' values of the
Galactic constants are in error at this level of precision.

More detailed fits to the rotation curve face the additional problem
that the shape of this curve becomes rather uncertain in the innermost
parts of the Galaxy, where one might hope to be able to see the effects
of a central cusp.  In this region, the Milky Way's bar means that
material in the disk is following strongly non-circular orbits, and it
is not trivial to allow for this effect in the analysis of kinematic
observations to obtain a definitive rotation curve.

These uncertainties leave plenty of ``wiggle room'' for those who
would like to hold on to the CDM paradigm.  By adopting a somewhat
non-standard rotation curve in its innermost parts, and by invoking a
modest redistribution of halo material out to larger radii, Klypin,
Zhao \& Somerville (2002) obtained a completely satisfactory match
between the observed rotation curve of the Milky Way and the
predictions of a multi-component model incorporating an initial CDM
cusp.  Thus, this test is not by any means conclusive.  Further
refinements in the data and modeling might in future allow a more
definitive result, but the large number of systematic errors that lurk
in the analysis mean that we are still a long way from such a
discriminating test.

\section{Halo Shape}\label{sec:shape}
One of the systematic effects that we have thus far ignored is that
the dark matter halo is unlikely to be exactly spherical.  If we allow
the halo to be flattened, then the simple argument from which we
derived a value of $\alpha \approx 0.4$ breaks down.  In particular, a
halo that is flattened towards the disk will concentrate the dark
matter closer to the Galactic plane, so that the average dark matter
density at a distance $R_0$ from the Galactic center will be less than
the value of $\rho^{\rm DM}(R_0, 0)$ that we measure in the Galactic
plane.  If we correct this over-estimate in the value that we have
used to normalize the dark matter density profile, then a larger value
of $\alpha$ is required to produce the same total mass of dark matter
interior to $R_0$.

\begin{figure}
\plotone{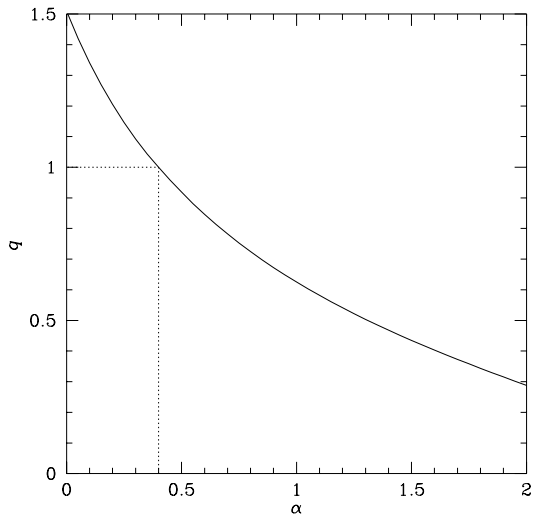}
\caption{Plot showing the locus of a family of power law axisymmetric
halos of index $\alpha$ and axis ratio $q$, all of which have the same
density at unit radius in the symmetry plane.  All these halos produce
the same gravitational field at unit radius as a spherical ($q = 1$)
halo with $\alpha = 0.4$ (highlighted with a dotted line).}
\label{fig:qalpha}
\end{figure}

The calculation is a little more complicated than this heuristic
argument would suggest, since flattening the dark matter distribution
also alters its gravitational potential.  Nonetheless, it is fairly
straightforward to calculate the acceleration due to gravity from
ellipsoidal power-law density distributions with different values of
$\alpha$ and degrees of flattening as parameterized by the
polar-to-equatorial axis ratio $q$.  One can show that all density
distributions with the same value of $\rho^{\rm DM}(R_0, 0)$ but
different values of $\alpha$ and $q$ will produce the same
gravitational acceleration in the Galactic plane at the Solar radius
if they have the same value for the quantity
\begin{equation}\label{eq:qalpha}
I(q, \alpha) = q\int_0^1 {x^{2 - \alpha}\,{\rm d}x \over \sqrt{1 - (1 - q^2)x^2}}
\end{equation}
[c.f.\ Binney \& Tremaine (1987) equation (2-91)].  This integral
equation can be solved by a simple iteration to obtain pairs of values
of $\alpha$ and $q$ that produce the same value of $I$.
Figure~\ref{fig:qalpha} shows the family of parameter values that
produces identical values of $I$ -- and hence the same acceleration in
the Solar neighborhood -- as the spherical $\alpha = 0.4$ halo derived
in Section~\ref{sec:cusp}.  Clearly, if we allow arbitrary flattening
of the halo, we can obtain a wide range of power law indices,
including the value of $\alpha \approx 1$ favored by the simplest CDM
models.  Perhaps more interestingly, we can turn this calculation
around.  If it is assumed that the dark matter is distributed in an
$\alpha \sim 1$ cusp, as the CDM models require, then
Figure~\ref{fig:qalpha} shows that the halo has to be modestly
flattened to an axis ratio of $q \sim 0.7$ in order to provide the
correct centripetal acceleration at the Solar radius.

This value can be compared with what is found through other lines of
investigation.  The difficulty in measuring the shape of the Milky
Way's halo by other means is that most probes of its potential only
sample the equatorial plane, so provide no direct handle on the mass
distribution in the polar direction.  However, some tracers are
affected by this third dimension. 

One of the strongest tests is provided by the variation in thickness
of the Galaxy's gas layer with radius.  Because the amount of mass
close to the plane decreases with radius, the local gravitational
field in the $z$ direction also decreases.  This lower field is less
effective at pulling the gas to the plane, so the thickness of the
layer increases with radius.  However, if the dark matter halo is
flattened, the dark matter is concentrated disproportionately toward
the Galactic plane, increasing the vertical gravitational field and
thus somewhat suppressing this tendency of the gas layer to flare with
radius.  Olling \& Merrifield (2000) compared the observed flaring of
the Galactic gas layer to the predicted thickness of the layer in
hydrostatic equilibrium in plausible mass models of the Milky Way with
a variety of halo flattenings.  This analysis showed that satisfactory
fits could be obtained with flattenings of $0.7 < q < 0.9$.

An exciting new probe of the third dimension is offered by the tidal
debris of merging remnants.  In the Milky Way, the detritus from the
Sagittarius Dwarf Galaxy has now been traced all around the sky
(Majewski 2003).  The fact that it forms a coherent stream over such a
large angle places a strong limit on the degree of halo flattening.
Only in a spherical potential would this orbit stay in a single plane;
in a flattened potential, the orbit would precess around, destroying
the coherence of the stream.  Ibata et al.\ (2001) used this reasoning
to argue that the carbon stars they had traced from the stream must be
orbiting in a potential that is close to spherical, with $q \ga 0.9$.
However, the case here is still not entirely clear-cut: if the carbon
stars that had been detected came from a stream that had wrapped
around the Milky Way several times, as the Ibata et al.\ model
suggested, then at any point around the stream there should be stars
on different ``laps'' around the Galaxy, with rather different
kinematics.  However, the observed kinematics of these carbon stars
shows a lot of coherence, suggesting that most stream members are on
the same lap.  If the stream is more localized in this way, then the
effects of precession would be reduced, and so a flatter halo could be
possible.  Clearly, more data are required.

\begin{figure}
\plotone{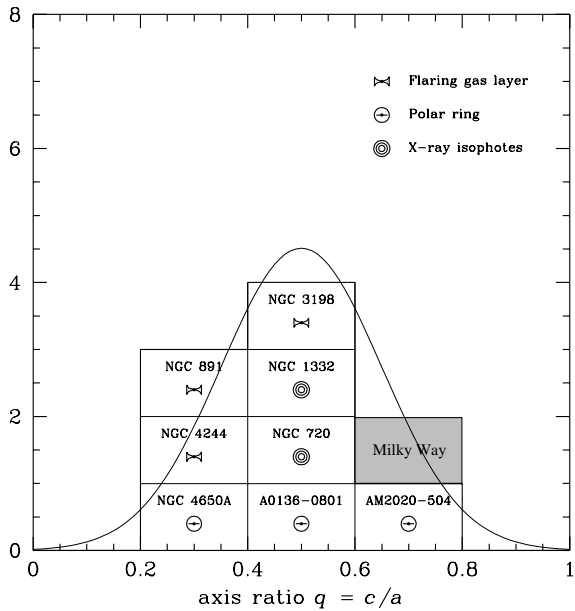}
\caption{Histogram showing how the Milky Way fits in with the
distribution of halo shapes for the handful of galaxies where
reasonably reliable measurements have been made.  The line shows the
predicted distribution for a CDM cosmology from Dubinski (1994).}
\label{fig:qdist}
\end{figure}

Another test of the plausibility of a flattening of $q \sim 0.7$ is by
comparison with the flattenings of the halos in other galaxies.  Such
measurements are equally difficult in other systems, but there are a
few objects with reliable measurements, based either on the gas layer
flaring method, or observations of polar rings (which are similar in
nature to the merger stream discussed above), or the shape of the
isophotes of X-ray emission from hot gas [which provides a rather
direct measure of the shape of the gravitational potential in
early-type galaxies -- see Buote (2003)].  The Milky Way is placed in
the context of these measurements in Figure~\ref{fig:qdist}.  The
amount of data is depressingly small, but it is already apparent that,
although the Milky Way's halo is a little rounder than average, it is
by no means exceptional.

Figure~\ref{fig:qdist} also shows the approximate distribution of halo
flattenings that were produced in Dubinski's (1994) CDM simulations.
Using this distribution as a test of the theory is rather better
than looking at the steepness of any central cusp in the dark matter
distribution: as discussed in Section~\ref{sec:cusp}, there are a
number of ways of altering the primordial degree of cuspiness at the
centers of halos, but the larger-scale shape of the halo is a lot
harder to mutate.  As is apparent from Figure~\ref{fig:qdist}, CDM
passes this test with some ease, although the shortage of data limits
the scope for detailed comparison.

\section{Conclusions}
Astronomers often proceed dangerously rapidly from saying ``this
theory is completely implausible'' to asserting ``I always said that
this theory was right, and, by the way, I invented it.''  CDM is no
exception to this overly-rapid process, so it is certainly still worth
testing the theory critically wherever possible.  In this regard, the
unique {\it in situ} observations that one can make of the dark matter
density in the Milky Way offer important tests of the theory.

Happily (or unhappily, depending on your perspective), the Milky Way
seems to pass these tests fairly easily.  The apparent discrepancy
between the predicted density cusp of a CDM halo and the observed mass
of dark matter in the inner Milky Way can be explained astronomically
(e.g., through errors in the adopted rotation curve) or
astrophysically (e.g., via redistribution of the primordial halo by
a bar).  One can even invoke CDM itself to explain the discrepancy: the
theory generically predicts that halos should be flattened, and when
such flattening is introduced in the modeling a steeper cusp is
inferred for the Milky Way.

Other measures of halo shape are also more-or-less consistent with
this picture.  The one potentially conflicting result comes from the
study of the Sagittarius Stream: if future work shows that this stream
really is wrapped several times around the Galaxy without significant
precession, then the Milky Way must have a halo that is very close to
spherical, which would be rather uncomfortable for CDM.  However, even
such a strong result would not be a killer blow against CDM, since the
theory predicts a wide range of halo shapes, a few of which will be
close to spherical.  

Indeed, this shortcoming illustrates what is probably the fundamental
limiting factor in using the Milky Way as a laboratory to test CDM.
The theory is sufficiently flexible that any property of the Milky Way
is likely to lie somewhere within the range of possibility.  The real
test must surely come from comparing the properties of large samples
of galaxies to the predicted distribution of the population in CDM
cosmology.  If, for example, star streams in other galaxies were all
found to require spherical halos, then CDM would be in deep trouble.

\end{document}